\documentstyle[11pt,epsfig]{article}

\newcommand{\be}{\begin{equation}}
\newcommand{\ee}{\end{equation}}
\newcommand{\bea}{\begin{eqnarray}}
\newcommand{\eea}{\end{eqnarray}}

\newcommand{\SG}{\sigma}
\newcommand{\al}{\alpha}
\newcommand{\ep}{\varepsilon}
\newcommand{\nn}{\nonumber}

\begin{document}

\begin{center}{ \Huge {\bf Running coupling and fermion mass 
in strong coupling $QED_{3+1}$}}
\end{center}
\
\\
\
\
\begin{center} {\large \bf Vladim\'{\i}r \v{S}auli}
\end{center}
\
\

\begin{center}{ \it Department of Theoretical Physics,
Nuclear Physics Institute,}
\end{center}
\begin{center}{\it \v{R}e\v{z} near Prague, CZ-25068,
Czech Republic}
\end{center}

\

\begin{center} {\large \bf Abstract}
\end{center}

Simple toy model is used in order to exhibit the technique of 
extracting the non-perturbative  information about 
Green's functions  in Minkowski space. 
The effective charge and the dynamical electron mass are
calculated  in strong coupling 3+1 QED by
solving the coupled Dyson-Schwinger equations for electron and
photon  propagators. The minimal Ball-Chiu
vertex   was used for simplicity and we impose the Landau gauge fixing
on QED action.  The solution  obtained separately   in  Euclidean and
Minkowski space were compared, the  latter  one was extracted with the
help of spectral technique.

\newpage

\section{Introduction}

In quantum field theory and even in physics at all, the dispersion relations (DRs)
were recognized as providing useful connections between physical quantities.   

Very recently \cite{ALDEFIMA2003}, it was recognized that 
in the Quantum Chromodynamics (QCD) in Landau gauge
the full (non-perturbative)  gluon propagator is (very likely) an analytic function in the 
whole complex  plane of momenta except the positive-timelike \cite{CONV}
real $p^2$ half-axis. This has been achieved 
by the analytical  fits of 
some recent solutions of Dyson-Schwinger equations (DSEs) \cite{FISCHER} 
and by analytical parameterization of  some contemporary lattice data.
The reasonable spacelike domain agreement of these fits with the recent Euclidean 
data  give us a good guidance on the possible analytical structure of the Green's function 
in timelike axis of momenta.   
The similar was argued for the quark propagator, however in this case the observed singularities
do not occur exactly on - but rather say - very close to 
the real positive  $p^2$ half-axis.

 In this paper, we instead to make an analytical guesses of already known Euclidean results 
we start with the assumption of $C-R_+$ analyticity of the
Green's functions which allows us to solve the DSEs system directly in Minkowski space.
It is well known  that such an assumption is intimately related with the
formulation of the appropriate dispersion relations (DRs) and spectral
representation (SR) of Green's functions (for more then the 
derivation of DRs in the context of QCD  see \cite{KON2003}). 
The main advantage of such an approach is the possibility to the
knowledge of the propagators at the whole range of momentum $p^2$ 
while the main inconvenience of this method is 
the  necessity of  'inhuman' effort when  the absorptive parts 
of Green's functions are actually derived. 
Unfortunately at the contemporary  stage of our calculation we are not able to write 
down the appropriate DRs for ghosts and transverse gluons.
Due to this  fact we confine ourself to the less complicated 
(at least from the technical point of view) case of 
one flavor strong coupling Quantum  Electrodynamics
and we leave the   involvement of Yang-Mills theory  for the future study.

In the present
paper and in contrast to \cite{SAULI} we adopt the  Ball-Chiu (BC) vertex 
\cite{BALLCHIU} which is the known minimal ansatz consistent with 
the Ward-Takahashi identity (WTI).  Implementing this into the equations for electron and
photon we then solve the corresponding DSE also for the photon polarization function.
In this sense, the presented study is the extension of the previous  numerical study of 
the renormalized electron mass in the strong coupling
QED in simplest approximation  to the DSEs: 
the bare vertex and bare photon propagators were employed \cite{SAULI}.

Up to the case of  perturbative theory the direct Minkowski space treatment of
DSEs is usually rather involved and the progress is adequately less when 
compared with the relatively large amount of calculations performed 
in Euclidean space (for a list of references see the paper \cite{SAULI} 
and also the paper \cite{BICUDO} where the approximative analytical solutions are discussed)  . 
Although the spectral approach is rapidly getting technically rather
involved when one goes beyond lowest order truncation, one of the
purpose of this paper is to demonstrate that it is still
manageable for the considered model with WTI respecting  vertex. As in the paper \cite{SAULI}
we again compare in detail the solutions for fermion propagator in
Minkowski and Euclidean space, besides we compare for the first
time also the photon propagator. 
There is a another important reason  for the use of improved  vertex.
It is long time known that the scalar part of BC vertex dramatically affect the analytical structure
of the DSEs \cite{BURDEN} (which paper is the extended study  of the
Munczek-Nemirovsky model \cite{MUNNEM}), for the earlier study of the 
quark gap equation in the axial gauge see \cite{MUNCZEK}. 
In this place we should stress that the recent study of QCD DSEs
confirms this suggestion: {\it the bare vertex leads 
to the complex singularity of the chiral limit quark propagator} . 
The authors of Ref. \cite{ALDEFIMA2003} observed that  the scalar part of the BC  vertex 
in used plays the crucial role in the analytical structure of 
quark propagator and one can conclude
that the inclusion of {\it the scalar part of the BC  vertex 
leads again to the real singularity
of quark propagator}, noting that the same is true for quenched 
QED in the chiral limit 
(note only that  the full Curtis-Pennington vertex \cite{CURTIS} was used 
which fact has no large significance  due to the Landau gauge employed).

Contrary to QCD, the considered model here 
is not an asymptotic free theory and posses the additional complication due to the 
triviality statement. It  requires the
introduction of ultraviolet cut-off function $f(\Lambda)$ which  well known
fact is clearly confirmed by our numerical analysis.
However, we  stress here that QED is
trivial and the appropriate solution of DSEs (Euclidean and spectral
as well as) collapse due to the presence of Landau singularity
{\cite{LANDAU}}, we regard the strong QED as an useful toy model which is clearly
reliable when we take cutoff reasonably smaller then the position of the  expected
 of   Landau singularity.  Further, in order to fully 
specify our model we  consider non-zero  bare electron mass $m_0$.

For comparison we have also calculated the propagators in unquenched
approximation (i.e., $\Pi(q^2) \neq 0$), but with the bare vertex. In
that case the vacuum polarization tensor is not gauge invariant and
contains two independent scalar functions. We impose the transversality
by hands before the numerical solution. As expected  from the earlier studies
\cite{WILLIAMS},\cite{BLOCH} the
results obtained in Landau gauge are very close to the results
calculated with BC vertex. This pleasant but extraordinary property of
Landau gauge fixing makes the  solutions of DSEs with the bare vertex
approximation meaningful. However the similar conclusion was made in some QCD studies
 \cite{BENDER}, \cite{ROBERTEK} one should be aware about the incidence
on the analytical structure of the quark propagator.
Although when using the bare vertex 
the  transverse projection by hands is not theoretically justified 
and although the analyticity assumptions seems not to be fully justified 
the usage of bare gauge vertices  in Landau-like gauges remains popular in 
studies of more complicated gauge models
(like extended, walking Technicolor etc.  for review see \cite{KING} ).
This is why we believe that
showing in detail the effects due to BC vertex is interesting: in our
studies its effects are less than 10 per cents even for rather large
coupling constant. The strong coupling QED should serve as an
instructive tool for its 'simplicity' and the proposed technique could
be helpful elsewhere.

The layout of the article is  following: 
In  next section we review the DSE formalism
and describe the model. 
The section 3 is devoted to  the solution of DESs in  Euclidean space. 
In the section 4 the DSEs are written in Minkowski
space and the desired DRs for electron selfenergy 
and gluon polarization function are derived.  The  numerical results 
are presented in the section 5 and then we summarize.

\section{Model- Unquenched QED with BC vertex}

In  this section we  review the basic elements of the
considered model and set the notation and conventions used in the main
part of the paper. 
First of all, let us  stress some differences and improvements used in this paper when compared
with until now published works  that are dealing with QED DSEs.   
Using the full photon propagator, instead of the bare propagator that authors of the papers
\cite{HAWES},\cite{HAWES1}, \cite{WILLIAMS} (bare and improved vertices, Euclidean space), \cite{SAULI} 
(bare vertex, Minkowski space) used is an important improvement.
Further we use  the form of the vertex that is consistent with WTI instead of bare vertex
$\gamma_{\mu}$ that was used in the study of renormalized DSEs in unquenched QED \cite{RAKOW}.
This leads to the solution which reflects not only the effect of running coupling caused by fermion loop
but also to the vacuum polarization which is automatically  transverse. 
Thus we can call our solution as a gauge covariant one since it naturally respect the conservation low
that follows from the gauge invariance of QED Lagrangian.   
The three lowest DSEs read:
\bea
\left[S\right]^{-1}= \left[S_0\right]^{-1}-\Sigma \quad &;& \quad
\Sigma=e^2\int S\Gamma^{\mu}G_{\mu\nu}\gamma^{\nu} \, ,
\label{DSEferm}\\
\left[G_{\mu\nu}\right]^{-1}=\left[G_{0\mu\nu}\right]^{-1}-\Pi_{\mu\nu}
\quad &;&
 \Pi_{\mu\nu}=  e^2\, Tr\,\int  S\Gamma^{\mu}S\gamma^{\nu} \, ,
\label{DSEphot}\\
\Gamma^{\mu}=\gamma^{\nu}+e\int S\Gamma^{\mu}S {\cal M}
&=&\Gamma^{\mu}_{L}+\Gamma^{\mu}_T \, ,
\label{DSEvert}
\eea
where ${\cal M}$ is the electron-positron scattering kernel (without annihilation channel),
 $S$ is the full fermion propagator
\be
\label{parametr}
S(p)=\frac{1}{A(p^2)\not p-B(p^2)}=\frac{F(p^2)}{\not p -M(p^2)} \, ,
\ee
parametrized in a usual way in terms of two scalar functions  $A,B$.
Equation (\ref{DSEferm}) then reduces to a coupled set of two scalar
equations for the functions $A,B$ or equivalently for the dynamical
mass $M=B/A$ and the electron renormalization function $F=A^{-1}$.
When the interaction is neglected, Eq.\ (\ref{parametr}) reduces to the
free propagator: $ S_0^{-1}=\not p- m_0$, $m_0 $ being the bare
electron mass. The sixteen Lorentz components of Eq.\ (\ref{DSEphot})
can  be reduced to single equation for  polarization function  $\Pi$
\be
\label{scalarpol}
\Pi^{\mu\nu}(q)={\cal P}_T^{\mu\nu}q^2\Pi(q^2)\quad ; \quad
{\cal P}^{\mu\nu}_T=(g^{\mu\nu}-\frac{q^{\mu}q^{\nu}}{q^2}) \, ,
\ee
by virtue of the gauge invariance $q^{\mu}\Pi_{\mu\nu}=0$. The function
$G_0^{\mu\nu} $ is the quenched approximation ($\Pi=0$) to the full photon
propagator $G^{\mu\nu}(q)$, which is purely transverse in Landau gauge,
\be
G^{\mu\nu}(q)=-\frac{{\cal P}^{\mu\nu}_T}{q^2\left[1-\Pi(q^2)\right]}
\, .
\ee

The functions $\Gamma^{\mu}_{L}$ and $\Gamma^{\mu}_T$ in
(\ref{DSEvert}) are the longitudinal and transverse parts of the full
vertex $\Gamma^{\mu}$. Multiplying the vertex by the photon momentum $p-l$
one gets the WT identity
\be \label{WTI} S^{-1}(p)-S^{-1}(l)=(p-l)_{\mu}\Gamma_L^{\mu}(p,l) \, ,
\ee
while from its definition $\Gamma^{\mu}_T(p,l).(p-l)=0 $.
Any truncation  of DSEs system  leading to gauge covariant solution of DSEs must involve
vertex satisfying WTI. Instead of solving own equation for $\Gamma$ there exists much economic way.
Within the requirement of right Lorentz transformation property, charge conservation
and the unique limit when $p\rightarrow l$ (the absence of 
kinematic singularity) the longitudinal part 
$\Gamma^{\mu}_{L}$:
\bea
\label{gamaL}
\Gamma^{\mu}_{L}(p,l)&=&\frac{\gamma^{\mu}}{2}\left(
A(p^2)+A(l^2)\right) \nn \\
&+&\frac{1}{2}\frac{(\not p +\not l)(p^{\mu}+l^{\mu})}{p^2-l^2}
\left( A(p^2)-A(l^2)\right) -\frac{p^{\mu}+l^{\mu}}{p^2-l^2}
\left( B(p^2)-B(l^2)\right) \, .
\eea
was found in the paper \cite{BALLCHIU} while the transverse part  $\Gamma^{\mu}_T(p,l)$
can be decomposed into the eight component vector basis (see for instance 
\cite{BALLCHIU} or\cite{PENNINGTON} for the details)
\bea
\Gamma_T^{\mu}(p,l)=\sum_{i=1}^8t_iT_i^{\mu}.
\label{covform}
\eea
with coefficient functions $t_i$ unspecified in general.

The one loop analysis and determination of  $t_i$ was performed 
in the paper \cite{PENNINGTON}
in arbitrary covariant gauges. The two loop results was obtained in the Feynman 
gauge also \cite{BALLCHIU}. 
The minimal version of the gauge covariant vertex - BC vertex simply neglect 
the transverse  part and hence is given by the Eq. (\ref{gamaL}). 
Some better improvement of the full vertex $\Gamma$  necessarily 
differs only by its transverse part $\Gamma_T$. In chiral symmetric case with
given truncation of DSEs and within the requirement of multiplicative renormalizability 
the additional constraints  on the transverse pieces of the vertex were found \cite{DONG}.
Also the additional information was obtained from the requirement  gauge independence of 
chiral symmetry breaking in quenched QED \cite{BASHIR}.
Although, as it follows form the above two notations 
the BC vertex has not probably most ideal form for some 'definitely conclusive' 
nonperturbative study of QED, nevertheless
we use this vertex for its simplicity. The  implementation of the methods 
used in  \cite{DONG}, \cite{BASHIR}
and mainly  their true Minkowski space extension is far away from 
triviality and thus remains the challenging task for future investigation.

In the Landau gauge and  in the ladder approximation of the  electron
DSE  there is no self-energy  contribution to the  electron
renormalization function since the   Feynman (F) pole part  of  photon
propagator $G_{\mu\nu}$ is exactly canceled by  the contribution of
longitudinal (LO) $q_{\mu}q_{\nu}$ part of $G_{\mu\nu}$.  Explicitly
their absorptive parts satisfy:
\be
\Im A_F(\omega)=-\Im A_{L0}(\omega)=\frac{e^2 m}{(4\pi)^2}
\int d a \left(1-\frac{a^2}{\omega^2}\right)\sigma_v(a)\Theta(\omega-a) \, ,
\ee
which leads to $A=A_F+A_{LO}=1$ (here $\sigma_v(a)$ is  Dirac
coefficient Lehmann function (\ref{lemani}), $m$ is the physical (pole)
electron mass). It is also well known that the property $A\simeq 1$
persists beyond the bare vertex approximation and that in angle
approximation \cite{KONDO} $A=1$ is even valid exactly. The above mentioned study
\cite{RAKOW} of unquenched QED shows rather small and irrelevant violations from the 
identity $A=1$ which then  turns to  few percentage error  only when momentum approaches  
the ultraviolet cutoff $\lambda$ (note, the  condition  $A(0)=1$ was exactly imposed in the paper
\cite{RAKOW}). Further study on 
dynamical mass generation in  unquenched supercritical QED \cite{BLOCH},
\cite{KONDO} confirm this also in this case. They  justify our approximation with at most ten
percentage deviation in the infrared region (when $A(\Lambda)=1 $ condition was imposed by the authors.
 To  reduce the complexity of our DSEs we explore this nice Landau gauge property 
and we put explicitly $A=1$ for all momenta. We should stress here that this neglection has no
effect on the gauge invariance  of polarization tensor, i.e. we still have 

\be
\Pi(q)^{\mu\nu}q_{\mu}=0
\ee
since the WTI (\ref{WTI}) is not violated.

When we renormalize we adopt the standard notation for the renormalization constants $Z_1$,
$Z_2$ and $Z_3$ (see e.g.,\ \cite{ZUBER}). From the approximation
employed it follows that $Z_1=Z_2=1$ which is in agreement with  the
multiplicative renormalizability and WTI. Furthermore, the unrenormalized vacuum polarization
$\Pi(\mu)$ should be  absorbed into the renormalization constant $Z_3$.
Similarly,  the unrenormalized electron  
self-energy $Tr\, \Sigma(\mu)/4$ is absorbed into the constant $Z_m$.

\section{Solution of DSEs in Euclidean space}

The DSEs are often solved in Euclidean space after the Wick rotation
$k_0 \rightarrow k_{1E}$ is made for each momentum. Then the loop
integrals should be free of singularities and the Green functions are
found for positive Euclidean momentum $k_E^2=k_1^2+k_2^2+k_3^2+k_4^2$.
If there is no additional singularity in the complex plane of momenta
(that would prohibit the validity of the naive  Wick rotation),  then
from the solution for some generic function $f(p^2_E)$  one would get
the  solution for Minkowski spacelike momentum   $f(p^2_M);\, p^2<0$.
The solution for timelike momentum would be in principle obtained by
the  analytical continuation of $f$ to the real axis $p^2_M>0$. In our
case, because of  QED triviality  in four dimensions, the assumption of
non-singular behavior holds only with the presence of UV cut-off. We
presume here, that  any numerical  attempt  to avoid  UV cut-off
implementation would lead to an  uncontrolled behavior of the
Gell-Mann-Low effective charge. The presence of the ultraviolet cut-off
is required not only due to the inner consistence (the Wick rotation)
but also   to ensure the numerical stability of our  calculation.

Substituting the  Ball-Chiu vertex into the DSEs and employing the
projection proposed in \cite{PENN2} (and successfully used  in
the papers \cite{BLOCH},\cite{FISCHER}) we obtain the following coupled
DSEs to be solved numerically:
\bea  \label{bloch1}
\Pi_U(x)&=&\frac{2\alpha}{3x\pi^2}\int d y \frac{y}{y+M^2(y)}\int d\theta \sin^2\theta
\frac{1}{z+M^2(z)}\biggr[\left(2y-8y\cos^2\theta+6\sqrt{yx}\cos\theta\right)
\nn \\
&+&
\frac{(M(y)-M(z))}{y-z}\left(M(y)+M(z)\right)
\left(2y-8y\cos^2\theta+3\sqrt{yx}\cos\theta\right) \nn\\
&+&3M(y)(M(y)-M(z))\biggl] \, , \\
 \label{bloch2}
M(x)&=&m_0+\frac{\alpha}{2\pi^2}\int d y \frac{y}{y+M^2(y)}
\int d\theta \sin^2\theta \nn\\
&&\frac{1}{z(1-\Pi(z))}
\left[3M(y)-\frac{(M(y)-M(z))}{y-z}\frac{2yx\sin^2\theta}{z}\right]\quad,
\eea
where $\alpha=e^2/4\pi$ and variables $x,y,z$ represent squares of
Euclidean momenta, $z=x+y-2\sqrt{yx}\cos\theta$. For details of the
derivation of (\ref{bloch1}),(\ref{bloch2}) we refer to refs.\
\cite{PENN2},\cite{BLOCH}.

The gauge invariance $q^{\mu}\Pi_{\mu\nu}=0$ implies that the photon
polarization tensor has to be of the form (\ref{scalarpol}.

In the DSE's formalism the gauge invariance of $\Pi_{\mu\nu}$  follows
from the gauge covariance of the Ball-Chiu vertex. This fact helped the
authors of \cite{PENN2} to construct simple recipe how to avoid
numerical  quadratical divergence which would be otherwise presented in
equation for $\Pi$. Anticipate here that it is convenient reduce the
photon polarization to a single scalar function also in our Minkowski
calculation, although there it is not a numerical necessity, but merely
matter of technical convenience.

The first line inside the brackets $[\dots ]$ of the Eq.\
(\ref{bloch1}) and the first term in the brackets $[\dots ]$ of Eq.\
(\ref{bloch2}) represent the kernels of the bare vertex approximation
(with only $\gamma_{\mu}$ retained). They give  dominant contributions
to the dynamical mass of electron and the vacuum polarization as well.
Neglecting the vacuum polarization effect (putting $\Pi(z)=0$) in the
Eq.\ (\ref{bloch1}), the  equation for $M$ can be further simplified.
This so-called ladder approximation of fermion DSE is represented by
one dimensional momentum integral equation first derived in ref.\
\cite{JOHNSON} and used  also in the Euclidean  confinement study
\cite{KUGO}.

\section{Direct treatment in  Minkowski space}

Assuming analyticity for complex $p^2$ in  $C-R_+$ Gaussian plain with indicated cut
and using the known asymptotic  behavior of the propagator one can derive the
appropriate   Lehmann representation (LR) for the propagators ( without positivity).
 The appropriate LR for the fermion propagator in parity conserving theory reads
\be
\label{lemani}
S(\not\!p)=\int\limits_{0}^{\infty} d\omega \frac{\not\!p\SG_v(\omega)+
\SG_s(\omega)}{p^2-\omega+i\epsilon}=
\frac{r}{\not\!p-m}+
\int d\omega \frac{\not\!p\SG_{v(c)}(\omega)+\SG_{s(c)}(\omega)}
{p^2-\omega+i\epsilon} \, ,
\ee
where we integrated out the single particle state contribution to  the
full Lehmann weight $\SG^{(1.p.s.)}(a)=r\delta(a-m^2)$. The remaining
term $\SG_{(c)}$ is assumed  to be a real and continuous spectral
density that originates from  the interaction. Similarly  we can write
for the photon propagator in linear covariant gauges
\be \label{fotoni}
G^{\mu\nu}(q)=\int\limits_{0}^{\infty} db
\frac{\sigma_{\gamma}(b)\left(-g_{\mu\nu}+\frac{q^{\mu}q^{\nu}}{q^2}\right)}
{q^2-b+i\epsilon}-\xi \frac{q^{\mu}q^{\nu}}{q^2}\:,
\ee
where the single photon spectrum $r_P\delta(b)$ can be integrated out
as in the previous case.

Due to the asymptotic the   dispersion  formula for  fermion
mass function $B$ requires one subtraction
\bea   \label{selfcon}
B(p^2)&=&m(\mu)+\int d\alpha
\frac{\rho_s(\alpha)(p^2-\mu^2)}{(p^2-\alpha+i\ep)(\alpha-\mu^2)}\, ,
\eea
where  $m(\mu)$ is the renormalized mass at the scale $\mu$. Similar
relation could be derived  for the function $A$, but here $A=1$ and
therefore  the function $B$  represents renormgroup mass function.

Renormalized photon polarization function
in momentum subtraction scheme reads:
\be  \label{mom}
\Pi_R(q^2,\mu'^2)=\int\limits_{0}^{\infty}
d\omega \frac{\rho(\omega)(q^2-\mu'^2)}{(q^2-\omega+i\epsilon)(\omega-\mu'^2)} \, ,
\ee
where we distinguish two possibly different renormalization scales $\mu,\mu'$.
The appropriate renormalization accompanied by the detailed derivation
of  dispersion relations (\ref{selfcon}) and (\ref{mom}) is given
in the next two sections. 

In the perturbation theory the relations for $\rho$ is usually represented by the series
expanded in the coupling constant. Here the appropriate relations for $\rho$ and $\rho_s$
are represented by the integral equations involving the Lehmann functions $\sigma$'s and even the 
function $\rho$ itself. Together with two additional
equations for the $\sigma's$  they form the set of the  so called {\it Unitary Equations} (UEs) 
(since these are the relations between the imaginary and real parts of propagator 
(and inverse of propagator) functions, here  we follow the paper \cite{SAULI} ).
In order to derive the UEs recall  the  
well known functional identity for distributions
\be
\frac{1}{x'-x+i\ep}=P \cdot \frac{1}{x'-x}-i\pi\delta(x'-x) \, ,
\ee
where $P \cdot$ stands for principal value integration. Making use of
the LR for $G^{\alpha\beta}$ and of the appropriate DR for $\Pi$ in
$G^{-1}$ and evaluating  the imaginary part of the unit tensor
$G_{\alpha\beta}^{-1}G^{\beta\gamma}$ one arrives  to the integral
equation:
\be  \label{eqforphot}
\sigma_{\gamma(c)}(a)=\frac{r_{\gamma}}{a}\rho(a)+[\sigma_{\gamma(c)}* \rho](a) \, ,
\ee
where we have adopted a shorthand notation for  real functional:
\be \label{principial}
[\sigma * \rho](a)=P \cdot \int_{T}^{\infty} dx\frac{\rho(a)\sigma(x)+
\sigma(a)\rho(x)\frac{a-\mu^2}{x-\mu^2}}{a-x}\, .
\ee
Notice that the both (\ref{eqforphot}) and (\ref{principial})  are
non-zero only for timelike (square of) momentum $a$ (here $a>T=4m^2$).

For the electron Lehmann weights we get in a similar way:
\bea  \label{Tak2}
\SG_{v(c)}(\omega)= \frac{f_1+m(\mu)f_2}{\omega-m^2(\mu)}
\quad &;& \quad
\SG_{s(c)}(\omega)=\frac{m(\mu)f_1+\omega f_2}{\omega-m^2(\mu)} \, ,
 \\
f_1 \equiv r\frac{m\rho_s(\omega)}{\omega-m^2}
 +[\SG_{s(c)}  * \rho_{s(c)}](\omega)
\quad &;& \quad
f_2\equiv r\frac{\rho_s(\omega)}{\omega-m^2}+
[\SG_{v(c)}  * \rho_s](\omega) \, ,
\eea
which are  non-zero only for the  timelike $\omega>m^2$.

The physical electron mass is defined by $S^{-1}(p=m)=0$ or
equivalently $M(m)=m$. Using the dispersion relation (\ref{selfcon})
the desired relation reads:
\be \label{masspole}
m=m(\mu)+\int dx\frac{\rho_s(x)(m^2-\mu^2)}{(m^2-x)(x-\mu^2)} \, .
\ee
The  residuum value $r$ of  pole part of the propagator  is fixed
already when the renormalization  procedure is done. 
Because of our numerical solution 
is is not convenient to determine $r$
by taking the on-shell limit $p\rightarrow m$ directly. 
The easiest way  to evaluate $r$ is the inspection
of  the real part of the identity $S^{-1}S=1$
evaluated at some arbitrary scale $p^2$. Choosing for instance $p=0$
one gets the desired relation
\be \label{residuum}
r=\frac{m}{m(\mu)-\int\frac{\rho_s(x)\mu^2}{x(x-\mu^2)}dx}-
\int\sigma_{s(c)}(x)\frac{m}{x}dx \, .
\ee
which helps us to avoid dealing with complicated infrared
singularities.

The original momentum space DSEs are now converted into a coupled set
of the UEs (\ref{eqforphot},\ref{Tak2})
complemented by the subsidiary conditions for  the residua and
thresholds (\ref{masspole}) and (\ref{residuum}).

To solve UEs one has to consider these six integral equations
simultaneously at all the positive  values of spectral variables where
the corresponding spectral functions are non-zero. Any internal
inconsistency (i.e.,\ spacelike  Green functions singularities
mentioned in the introduction) should be seen or felt when the UEs are
actually solved. The original momentum space Green functions are then
obtained through the dispersion relation for the proper function or
equivalently by the integration of  the spectral representation for the
full connected  propagators $S,G$. Checking  (numerically) this
equivalence verifies the internal consistence of the method.  Compared
to the Euclidean approach the  spectral approach has clear advantage of
``already known'' analytical continuation at all momenta. The
disadvantage of the spectral approach is its failure in the (confining)
regime where the underlying assumptions are not justified.

%PHOTONPPHOTONPHOTONPHOTONPHOTONPHOTONPHOTONPHOTONPHOTONPHOTONPHOTONPHOTON
\subsection{Photon propagator}

In a fixed  gauge the photon propagator is fully determined by the
gauge independent polarization function. We describe below the
derivation of once subtracted DR, following from the momentum space
subtraction procedure for photon polarization tensor. First we briefly
review the method in its perturbative context.

In $4+\ep$ dimensions and for spacelike momentum $q^2<0$ the one loop
polarization function can be written   as \cite{nevim}
\bea
\Pi(q^2)&=&\frac{4e^2}{3(4\pi)^2}\left\{\frac{2}{\ep}+\gamma_E-ln(4\pi)+
ln\left(\frac{m^2}{\mu_{t'H}^2}\right)\right.
\nn \\
&+&(1+2m^2/q^2)\sqrt{1-\frac{4m^2}{q^2}}
ln\left[\frac{1+\sqrt{1-\frac{4m^2}{q^2}}}{1-\sqrt{1-\frac{4m^2}{q^2}}}\right]
\nn \\
&-&\left.\frac{4m^2}{q^2}-\frac{5}{3}\right\}-\delta Z_3\mu_{t'H}^{-\ep}\, ,
\eea
where $\mu_{t'H}$ is t'Hooft dimensionfull scale. The  mass-shell
subtraction scheme defines $Z_3$ so that $\Pi^{MASS}_R(0)=0$ which
implies that the photon propagator behaves as free one near $q^2=0$.
Choosing $\delta Z_3$ to cancel entire $O(e^2)$ correction we find
\be
\delta Z_3^{MASS}=\lim_{q^2\rightarrow 0}\Pi(q^2)=
\frac{e^2}{12\pi^2}\left[\frac{2}{\ep}+\gamma_E-ln(4\pi)+
ln\left(\frac{m^2}{\mu_{t'H}^2}\right)\right]
\ee
and renormalized polarization function in mass-shell renormalization
prescription satisfies well known dispersion relation
\be
\label{drs}
\Pi_R^{MASS}(q^2)=\Pi(q^2)-\lim_{q^2\rightarrow0}\Pi(q^2)=\int\limits_{0}^{\infty}
d\omega \frac{q^2}{(q^2-\omega+i\epsilon)\omega}\, \rho(\omega)
\ee
with the absorptive part
\be
\label{oneloop}
\pi\rho(\omega)=\frac{\alpha_{QED}}{3}(1+2m^2/\omega)\sqrt{1-4m^2/\omega}
\, \Theta(\omega-4m^2) \, .
\ee
which is given in many standard textbooks (see for instance
\cite{BOGOLIUBOV}, where the result of the integration in (\ref{drs})
is  written also  for timelike momenta). Recall that the one loop
$\Pi_R^{MASS}$ represents also self-energy calculated in the  popular
$\overline{MS}$ scheme for the special choice of t'Hooft scale
$\mu_{t'H}=m$ \cite{SILVERS}). Finally, let us remind the   definition
of the off-shell momentum space subtraction:  $\delta Z_3=\Pi(\mu^2)$.
Making  redefinition of the electron charge  accompanied by the finite
subtraction of (\ref{drs}) we can immediately  write down  the desired
dispersion relation (\ref{mom}).

Now we turn our attention to the derivation of momentum space
subtracted $\Pi_R(p^2,\mu^2)$ with the dressed propagators and with the
full Ball-Chiu vertex included. As mentioned in the previous section
validity of  Ward-Takahashi identity for $\Gamma_{BC}$ naturally leads
to the transversality of the polarization tensor
\be
\label{cunik}
\Pi_{U,R}^{\mu\nu}(q)={\cal P}_T^{\mu\nu}q^2\Pi_{U,R}(q^2) \, ,
\ee
where $ {\cal P}^{\mu\nu}_T=(g^{\mu\nu}-\frac{q^{\mu}q_{\nu}}{q^2}) $
is the transverse projector and capital (R) indicates that renormalized
tensor (\ref{cunik}) must respect gauge symmetry of unrenormalized (U)
one.

The truly massless photons with $\Pi_R^{MASS}(0)=0$ are consequence of
the renormalization prescription
\bea
\label{onesub}
\Pi_R(q^2,0)&=&\Pi_U(q^2)-\Pi_U(0) \, ,
\nn \\
\Pi_U(q^2)&=&\frac{\Pi_U^{\mu\nu}(q)
\left[g_{\mu\nu}-C\frac{q_{\mu}q_{\nu}}{q^2}\right]}{3q^2} \,
\eea
with arbitrary constant $C$, applied on the full polarization tensor
\be
\label{tensor}
\Pi_U^{\mu\nu}(q) \equiv
ie^2 \int\frac {d^4l}{(2\pi)^4}
\, Tr \left[ \gamma^\mu \, S(l) \,
\Gamma^\nu(l,l-q) \, S(l-q) \right] \, ,
\ee
where the explicit dependence of Ball-Chiu vertex on fermionic momenta reads:
\be  \label{ball}
\Gamma^{\mu}_{L}(l,l-q)=
\gamma^{\mu}-\frac{(2l-q)^{\mu}}{(l-q)^2-l^2}[M((l-q)^2)-M(l^2)]\, .
\ee

As soon as we use WTI constrained vertex   the $C$ independence of
resulting $\Pi$ is evident but the right choice  of $C$ facilitates
derivation of DR. The reason is that the Pennington-Bloch \cite{PENN2}
projector
\be
{\cal P}_{\mu\nu}^{(d)}(q)=\frac{1}{d}
\left[g_{\mu\nu}-(d+1)\frac{q_\mu q_\mu}{q^2}\right]
\ee
cancels the contribution from $d+1$ space-time metric tensor
$g_{\mu\nu}$ which simplifies the actual calculations.

Let us now  derive  $\Pi_{R}$
\be
\label{receipt}
\Pi_{R}(q^2,\mu^2)=\frac{\Pi_{U}^{\alpha\beta}(q)
{\cal P}_{\alpha\beta}^{(3)}(q)}{3q^2}-
\Bigl\{q\rightarrow \mu\Bigl\}
\ee
for the case when only $\gamma^{\mu}$ part of $\Gamma_{L}^{\mu}$ is
retained. Substituting the  spectral representation (\ref{lemani}) into
the expression for  photon polarization function  (\ref{receipt}),
(\ref{tensor}) we immediately get
\bea
 \Pi_{U(\gamma_{\mu})}(q^2)&=&\frac{ie^2}{3q^2}\, Tr\,
\int\frac{d^4l}{(2\pi)^4} \int d\al \int d b \nonumber\\
&&
\frac{\left(\gamma_{\mu}-4\frac{q_\mu \not q}{q^2}\right)
\left[(\not l-\not q) \sigma_v(a)+\sigma_s(a)\right]
\left[\not l \sigma_v(b)+\sigma_s(b)\right]} {(l-q)^2-a)(l^2-b)}\, .
\eea
From now on we omit the spectral integrals and  assume that the
presence of any spectral function with  given arguments automatically
implies integration over these variables. Since we will include
explicitly  the boundaries (thresholds) in step functions in
the integral kernels,  all integrals can be taken from zero to
infinity $\int_0^{\infty}$. Moreover we label the measure $-i\,
d^4l/(2\pi)^4$ by $d_l$ and  we also suppress $i\epsilon$
factors in denominators.   Combining  the denominators with the help
of  Feynman parameterization then gives
\be
\Pi_{U(\gamma_{\mu})}(q^2)=8e^2\, \int{d_l}\int_0^1 dx
\frac{\sigma_v(a)\sigma_v(b)x(1-x)}
{(l^2+q^2x(1-x)-a x-b (1-x))^2} \, .
\ee
The remaining integral is logarithmic divergent. After the subtraction
\be \label{twotrem}
\Pi_{R(\gamma_{\mu})}(q^2;\mu^2)=\Pi_{U(\gamma_{\mu})}(q^2)-
\Pi_{U(\gamma_{\mu})}(\mu^2)
\ee
it leads to the finite dispersion relation. Although this procedure is
rather straightforward we present for completeness briefly intermediate
steps of the derivation.  The subtracting procedure (\ref{twotrem})
yields explicitly
\bea
\Pi_{R(\gamma_{\mu})}(q^2;\mu^2)&=&8e^2\int{d_l}\int_0^1 dx\int_0^1 dz
\nonumber\\
&&
\frac{ \sigma_v(a)\sigma_v(b)[x(1-x)]^2(q^2-\mu^2)(-2)}
{(l^2+(q^2-\mu^2)zx(1-x)+\mu^2x(1-x)-a x-b(1-x))^3}
\nn \\
&=&\frac{8e^2}{(4\pi)^2}\int_0^1 dx \int_{\frac{a x+b(1-x)}{x(1-x)}}^{\infty}d\omega
\frac{ \sigma_v(a)\sigma_v(b)x(1-x)(q^2-\mu^2)}
{(\omega-\mu^2)(q^2-\omega)}\:,
\eea
where (after the loop momentum integration) the substitution
$z\rightarrow \omega=\frac{a x+b(1-x)}{x(1-x)}+\mu^2-\frac{\mu^2}{z}$ was made.
Changing the order of integrations and integrating  over $x$
(the appropriate integrals are listed in the  Appendix A)
we obtain the DR:
\bea
\label{rojedna}
\Pi_{(\gamma_{\mu})}(q^2;\mu)&=& \frac{e^2}{12\pi^2}
\int\limits^{\infty}_{(\sqrt{a}+\sqrt{b})^2}
d\omega\frac{q^2-\mu^2}{(\omega-\mu^2)(q^{2}-\omega+i\epsilon)}
\nn \\
&&
\frac{\Delta^{1/2}(\omega,a,b)}{\omega}
\left[1+\frac{a+b}{\omega}-\frac{b-a}{\omega}\left(1+\frac{b-a}{\omega}\right)\right]
\sigma_v(a)\sigma_v(b)\:,
\eea
where $\Delta$ is the well-known triangle function
\be \label{triangle}
\Delta(x,y,z)=x^2+y^2+z^2-2xy-2xz-2yz \, .
\ee
Considering in the expression above for $\sigma_v(x)$ only the delta
function parts of spectral functions, i.e., $r_f\, \delta(x-m^2)$, we
just recover the one loop perturbative result (\ref{oneloop}) (up to
the  presence of electron propagator residuum $r_f$, which is assumed
to be close to $1$ when the coupling is small):
\bea
\Pi^{pole}_{R(\gamma_{\mu})}(q^2,0)= r_f^2\,
\frac{\al_{QED}(0)}{3\pi}\int\limits^{\infty}_{4m^2}
d\omega\frac{q^2}{\omega(q^{2}-\omega+i\epsilon)}
\, \sqrt{1-\frac{4m^2}{\omega}}\left(1+\frac{2m^2}{\omega}\right)\, .
\eea
We see immediately that $\Pi_R(0,0)=0$ as required and that using the
projector ${\cal P}^d$ naturally reproduces the perturbation theory in
its lowest order.

Using the prescription (\ref{receipt}) we now carry on the derivation
for the part of the  polarization function with the {\it remaining }
term  of Ball-Chiu vertex (second term in rhs of Eq.\ (\ref{ball})).
First we drop the  part of the vertex which is proportional to $l$
(because the photon propagator is transverse) and  take a trace which
leads to following  finite loop integral:
\bea
\label{holina}
\Pi_{U(rem)}(q^2)&=&-\frac{4e^2}{3q^2}\int d_l
\left[\frac{M(l)-M(l-q)}{l^2-(l-q)^2}\right] \nn \\
&& \hspace*{-2.0truecm}
\frac{ \left\{\sigma_v(a)\sigma_s(b)
\left[2l^2-8\frac{(l\cdot q)^2}{q^2}+3l\cdot q+3q^2\right]+
\sigma_v(b)\sigma_s(a)\left[2l^2-8\frac{(l\cdot q)^2}{q^2}-
3l\cdot q\right]\right\}}
{[(l-q)^2-a][l^2-b]}\, .
\eea
The next intermediate steps of the DR derivation are given in the
Appendix B, here we simply present the final result: The full polarization
function with the Ball-Chiu vertex satisfies the once subtracted DR
\be
\Pi_R(q^2,\mu^2)=\int\limits_{0}^{\infty} d\omega\,
 \frac{\left(\rho_{\gamma_{\mu}}(\omega)+
\rho_{rem}(\omega)\right)(q^2-\mu^2)}
{(q^2-\omega+i\epsilon)(\omega-\mu^2)}\, ,
\ee
where $\rho_{(\gamma_{\mu})}$ follows from (\ref{rojedna}) and
$\rho_{(rem)}$ from (\ref{holina}) and (\ref{dolina}). Explicitly, they
read:
\bea
\label{skutek}
\rho_{(\gamma_{\mu})}(\omega)&=&\frac{e^2}{12\pi^2}
\frac{\Delta^{1/2}(\omega,a,b)}{\omega}
\left[1+\frac{a+b}{\omega}-
\frac{b-a}{\omega}\left(1+\frac{b-a}{\omega}\right)\right] \, ,
\nn\\
&&
\sigma_v(a)\sigma_v(b)\Theta\left(\omega-(\sqrt{a}+\sqrt{b})^2\right)
\nn \\
\rho_{(rem)}(\omega)&=&\frac{e^2}{6\pi^2}
 \sigma_v(a)\sigma_s(b)\rho_S(c)
\left[\frac{F(\omega,c,c)-F(\omega,c,a)+F(\omega,b,a)-
F(\omega,b,c)}{(c-b)(c-a)}\right]
\, ,
\nn \\
F(\omega,c,a)&=&\frac{\Delta^{1/2}(\omega,c,a)}{\omega}
\left[\frac{a+c}{\omega}-2\frac{a-c}{\omega}
\left(1+\frac{a-c}{\omega}\right)\right]
\Theta\left(\omega-(\sqrt{a}+\sqrt{c})^2\right)\:.
\eea
These expressions in their full form have been used in  our numerical
calculation. No principal value integration is necessary and the whole
integrand has a regular limit when one spectral variable approaches
another. Recall that the ordinary integrals over the spectral variables
$a,b,c$ are implicitly assumed. The function $\pi\rho_{S(c)}$ is simply
$\Im M(\omega)$ and it is evaluated in the next section.

%ELECTRONELECTRONELECTRONELECTRONELECTRONEELECTRONELECTRONEL
\subsection{Fermion propagator}

In this section we show that the Ball-Chiu vertex
\be  \label{ballchiu} \Gamma^{\mu}_{L}(p-l,p)=
\gamma^{\mu}-\frac{(2p-l)^{\mu}}{(p-l)^2-p^2}[M((p-l)^2)-M(p^2)]
\ee
substituted to the  electron  self-energy
\be
\Sigma(p)=
e^2\int d_l \gamma^{\nu} S(p-l)\Gamma^{\mu}_{L}(p-l,p)G_{\mu\nu}(\xi=0,l)\:,
\ee
together with the assumed LR for  electron  (\ref{lemani}) and  photon
propagator (\ref{fotoni}) leads to the dispersion formula for the
dynamical mass (\ref{selfcon}). In $F=1$ approximation we can write
\bea
M(p^2)&=&m_o+\frac{Tr}{4}\Sigma(p)
\nn \\
&=& m_o+e^2 \frac{Tr}{4} \int d_l \gamma^{\nu}\frac{(\not\!p-\not l)
\sigma_v(a)+\sigma_s(a)}
{(p-l)^2-a}\Gamma^{\mu}_{L}(p-l,p)
\frac{-g_{\mu\nu}+\frac{l_{\mu}l_{\nu}}{l^2}}{l^2-b}\sigma_{\gamma}(b)
\, ,
\eea
where we have adopted conventions and notations of the previous
section. Relating bare mass and the  renormalized one by $m_o=Z_m(\mu)
m(\mu)$ and  absorbing  $\Sigma(p=\mu)$  into the mass renormalization
constant $Z_m(\mu)$ gives the finite mass $m(\mu)$ and  finite DR for
$\mu$ independent dynamical mass function  (\ref{selfcon}).

Let us again start with  the pure $\gamma_{\mu}$ matrix part of the
$\Gamma_L$ in $\Sigma$. It leads to the following DR:
\bea
\label{hmotny}
M_{(\gamma_{\mu})}(p^2)&=&\int_{m^2}^{\infty} d\omega \frac{p^2-\mu^2}{\omega-\mu^2}
\frac{\rho_{\gamma_{\mu}}(\omega)}{p^2-\omega+i\epsilon} \, ,
\nn \\
\rho_{(\gamma_{\mu})}(\omega)&=&-
3\left(\frac{e}{4\pi}\right)^2 \sigma_s(a)\sigma_{\gamma}(b)
\frac{\Delta^{1/2}(\omega,a,b)}{\omega} \, .
\eea
(The derivation is straightforward, see for  instance   Appendix of
Ref.\ \cite{SAULI}). This is a dominant momentum dependent part of $M$.

Using the {\it remainder} terms in Ball-Chiu vertex (\ref{ballchiu}) we
get the following contribution to $M$:
\be \label{linetwo}
M_{(rem)}(p)=2e^2 \int d_l
\frac{\left[\frac{(p.l)^2}{l^2}-p^2\right]\rho_S(o)\sigma_{\gamma}(b)\sigma_v(a)}
{[(p-l)^2-a][(p-l)^2-o][p^2-o][l^2-b]} \, ,
\ee
where we have self-consistently used the formula for difference of the dispersion
integrals for $M$:
\be \label{ayrtonsenna}
-\frac{(2p-l)^{\mu}}{(p-l)^2-p^2}[M((p-l)^2)-M(p^2)]=\int
d\gamma\frac{\rho_S(\gamma)(2p-l)^{\mu}}{(p^2-\gamma)((p-l)^2-\gamma)}\, .
\ee
Using the Feynman parameterization (\ref{linetwo}) is after some
algebra transformed into:
\be \label{bobr}
-2e^2\ \int_0^1 dx \int_0^1 dy\int d_l \,
\frac{\left[(p\cdot l)^2-l^2p^2\right]\rho_S(o)\sigma_{\gamma}(b)\sigma_v(a)}
{[(p-l)^2-ax-o(1-x)]^2[p^2-o][l^2-by]^2} \, .
\ee
Matching in (\ref{bobr}) two $l$-dependent denominators
(using a Feynman variables $z$), making a shift
$l=\tilde{l}+pz$ and integrating over the momentum $\tilde{l}$
yields the  result:
\be \label{easy}
\frac{3p^2e^2}{(4\pi)^2} \int _0^1 dx \int_0^1 dy \int_0^1 dz
\frac{\rho_S(o)\sigma_{\gamma}(b)\sigma_v(a)z(1-z)}
{[p^2z(1-z)-axz-o(1-x)z-by(1-z)][p^2-o]} \, ,
\ee
which is UV finite by construction.

It is now  easy to write down the DR following from (\ref{easy}) (Some
details of its derivation are given in the Appendix C). The 'dominant'
part following from the pure pole $\beta=0$ of the photon propagator
reads explicitly:
\bea \label{vysl}
\Im \frac{ M_{rem}^{pole}(\omega)}{\pi}&=&\frac{3e^2}{2(4\pi)^2}
P\cdot \int_{m^2}^{\infty}\frac{1}{\omega-u} \nn\\
&& \hspace*{-2.0truecm}
\left[\frac{u+a}{\omega^2}\sigma_v(a)\rho_S(u)\Theta\left(\omega-\frac{u+a}{2}\right)
+\frac{\omega+a}{u^2}\sigma_v(a)\rho_S(\omega)\Theta\left(u-\frac{\omega+a}{2}\right)\right]
\, .
\eea
To sum it up, the dynamical fermion mass is given by:
\be \label{completa}
M(p^2)=m(\mu)+M_{(\gamma_{\mu})}(p^2)+M_{rem}(p^2)\, .
\ee
Anticipating our numerical results: since the whole $M_{rem}$ changes the
numerical results only slightly (as compared to $M_{(\gamma_{\mu})}$),
we are approximating its imaginary part in our numerics just by the pole
contribution $M_{rem}^{pole}$.

\section{Numerical solutions and results}

First, let us describe some technical points of our numerical
treatment. First, consider the  spectral approach which is simpler from
the numerical point of view. After the formal derivation of the DRs for
electron self-energy and vacuum polarization function we introduce the
positive cut-off $\Lambda$ in the following way
\be
F(s)=\int_T^{\Lambda^2} dx\, \frac{s\rho(s)}{x(s-x+i\epsilon)}\, ,
\ee
where the function $F$ represents $\Pi$ or $\Sigma$. That is, the
absorptive parts of proper function $\pi\rho(s)$ is modified by a step
function $\rho(s)\rightarrow \rho(s)\Theta (\Lambda^2-s)$. The same
cut-off is then formally  introduced into the Lehmann representation
for propagators. With the cut-off implemented the set of equations that
have been numerically solved comprise: the coupled nonlinear integral
unitary equations (\ref{eqforphot}),(\ref{Tak2}) for photon and
electron Lehmann functions, the equation for absorptive part of
self-energies (\ref{skutek},\ref{hmotny},\ref{kick}) and necessary
conditions (\ref{masspole},\ref{residuum}). This set of equations is
solved by iterations. Then the various Green's functions are calculated
through the appropriate DRs and considered to be physical for
$|p^2|<\Lambda^2$.

We have found that no other approximation is necessary and the unitary
form of  DSEs converges under the iteration procedure, no matter
whether the bare or BC vertex is used. This procedure naturally fails
when the employed cut-off is rather close to the expected singularity
of the running constant. Before this numerics fails one observes only a
trace of this singularity -- the large growth of the effective charge
at $p^2$ close to $\Lambda^2$ Fig.~1). Then the dynamical mass appears
to be negative at large value of timelike momenta (see the curve for
$\Lambda^2=10^7$ and $\alpha(0)=0.4$ in  Fig.~4).

In our Euclidean treatment we adopted the  simplest cut-off functions:
The Heaviside step function $\Theta(-k_E^2+\Lambda^2)$ has been
introduced into the kernel of  DSEs (\ref{bloch1}),(\ref{bloch2}),
i.e., upper bound of integrals is replaced $\infty \rightarrow
\Lambda^2_E$. The value of the cut-off $\Lambda_E=10^7 M^2(0)$ is taken
to be exactly the same as in the spectral technique described above,
where the zero momentum electron mass $M(0)$ is used as a scale. After
the subtraction the DSEs for renormalized photon polarization $\Pi$ and
dynamical mass $M$ have been solved on suitable grid  $p_i^2 \in
(0,\Lambda^2)$. For the bare vertex the equations can be solved by
iteration without any numerical problem.  The form of the BC vertex
makes the numerical procedure more difficult, even in the quenched
approximation. For the unquenched solution this is even more
troublesome due to uncontrolled oscillations of the numerical
iterations of the running charge in the infrared region. Hence we
approximate the finite difference in (\ref{bloch1}),(\ref{bloch2}) by
the appropriate differentiation- (the similar trick was used in the
paper \cite{FISCHER}). Explicitly,  we replace:
\be \label{aproxim}
 \frac{(M(y)-M(z))}{y-z}\rightarrow M'(y) \, .
\ee
After making this approximation in the Euclidean DSEs with BC vertex we
were able to find the numerical solution for the coupling up to the
half of the critical coupling for  the bare vertex approximation.

The DSEs has to be renormalized with the help of subtraction (both in
bare vertex approximation and the BC vertex, modified as discussed
above). Since in this case the subtraction cannot be done analytically
(unlike in the Minkowski treatment), one has to implement it
numerically, which is not straightforward as we describe below. The
following expression for polarization function
\be
\Pi(x)=\int dx\, f(x,y)/x \, ,
\label{pigen}
\ee
need to be renormalized so that $\Pi_R(0)=0$ (for the explicit form of
the function $f$ see (\ref{bloch2})).  Doing this numerically with
reasonably high accuracy is not as simple task as for the spectral
approach or in a perturbation theory. We first solved the equation  for
the quantity
\be
\tilde{\Pi}(x)=\int dx \left[f(x,y)-f(0,y)\right]/x \, . \ee
After finding the solution for $\tilde{\Pi}$  we looked the limit
$x\rightarrow 0$ at which $\tilde{\Pi}(0) = K$ and subtracted this
constant in order to obtain 'right' $\Pi_R(x)=\tilde{\Pi}(x)-K$. When
one step is not sufficient we repeated the procedure. Without this we
were not able to find the Euclidean solution an accuracy comparable
with our Minkowski technique. For instance, for $\alpha=0.2$ making the
subtraction directly for (\ref{pigen}) leads to 35 per cent
underestimate of $\alpha(p^2)$ in the infrared region and 50 per cent
underestimate in the ``ultraviolet'' region. Compared to this, five
iterations of the procedure described above gives a satisfying with 0.2
per cent deviation in the infrared, the subtraction constants in
successive interpolations are  $K_i=+0.35,-0.12,+0.03,-0.008,+0.002$.
Furthermore, we use the linear interpolation and log (perturbative)
extrapolation to evaluate functions $\Pi(z)$ and $M(z)$ at
$z=x+y-2xy\cos{\theta}$ in Eqs (\ref{bloch1}),(\ref{bloch2}).

In the both formalisms we use one common renormalization scale
($\mu^2=0$) to define the  running coupling
\bea
\label{running}
\alpha(p^2)&=&\frac{\alpha}{1-\Pi_R(p^2)} \, , \nn \\
\alpha&=&\frac{e^2_R}{(4\pi)} \, ,
\eea
where we have explicitly used $\alpha=\alpha(0)$ and we omit explicit
dependence on  $\mu$ in R-label quantities for purpose of brevity. We
use the same scale in order to renormalize the electron mass. As a mass
scale of the theory we use $M^2(p^2=0)=1$ in arbitrary units.  The
momentum axis at all figures defined in this unit.

When the coupling $\alpha$ increases the pole mass and the residuum of
fermion propagator become different from their non-interacting values
($r=1 ; M_p=M(0)$). Couple of values of $r$ and $M_p$ following from
our solutions of DSEs with bare and BC vertex are shown in the Tab.1.
The residuum  $r$ is clearly renormalization (and gauge fixing) scheme 
dependent quantity. However on the physical ground one can expect that 
'one particle' contribution to an interacting
particle propagator is less than one, i.e. we would have naively 
$r<1$,  here the values of residua are greater 
then 1 which is  the  consequence of our subtraction scheme.
Not surprisingly, if the similar scheme and the same gauge are employed than 
the  property $r>1$ survive in  the case of quark propagator too
\cite{ALDEFIMA2003}.

\begin{center}
\small{\begin{tabular}{|c|c|c|c|} \hline \hline
$\alpha  $& 0.1 & 0.2 & 0.4  \\
\hline
$M_p /M(0)$ -BC& 1.044& 1.10 & 1.39 \\
\hline
$M_p /M(0)$ -BV& 1.042 & 1.09 & 1.23 \\
\hline
$r$ -BC &1.090& 1.22& 1.98 \\
\hline
r -BV& 1.085 & 1.19 & 1.53 \\
\hline\hline
\end{tabular}}
\end{center}
Tab.1 Pole mass and residua   of pole part of the electron
propagator. The label BC(BV) means the results calculated with BC
(bare) vertices. The coupling $\alpha$ is the value of running
charge at zero momenta.

Let us finally compare our numerical results obtained in the both
formalisms with bare or BC vertices.  The so-called photon
renormalization functions $G$ (it is defined by $G=\alpha(p^2)/\alpha$)
are compared for spacelike momenta in Fig.~1. One sees excellent
agreement between the solutions obtained in spectral and Euclidean
formalism. The one-loop perturbation theory (PT) result is added for
comparison. The correct pole mass $M_p$ is used in perturbative
formulas. The PT results always below the lines corresponding to the
DSEs solutions. This can be easily read from the bar vertex form of the
Euclidean DSE: the decreasing $M(x)$ enhances the function $|\Pi|$,
which being negative must enhance $\alpha(x)$. The same functions $G$
are displayed for timelike momenta in Fig.~2., where only the results
obtained from the unitary equations are presented. Again we would like
to stress that the differences between the solutions with  bare and BC
vertices are very small.

The expected exceptions are solutions with the value coupling constant
close to the one for  which the numerical solutions fail, that is
$\alpha_c \simeq 0.41$ and corresponding $G(p^2=-\Lambda^2)\simeq \Re
G(p^2=\Lambda^2) \simeq 2.6$. The dynamical fermion mass obtained from
DSEs is displayed for spacelike regime of momenta in Fig.~3.  The small
deviation of Euclidean results from the spectral ones can be explained
as a numerical error. The Fig.~4 shows the difference that follows from
the use of different vertices. The absorptive parts of $M$ for bare
vertex approximation are also displayed in this Fig. for the same
coupling $\alpha=0.1, 0.2, 0.4$. The negative damp of $\Re M$ is observed
for $\alpha=0.4$. At the end we should remind the reader that the bare vertex 
solution was already obtain by Rakow in the paper \cite{RAKOW}. 
The aforementioned approximative independence on the cutoff value 
was explicitly shown in this work. In order to check the consistence we 
took the cutoff to be as in the paper \cite{RAKOW}, for instance: 
$M(0)/\Lambda=10^{-3}$ together with  increasing of  the  renormalized coupling as
$ \alpha=0.5$  then we the agreement between us and  the Rakow solution
was found.

\section{Conclusion}

The Dyson-Schwinger equations for strong coupling QED were solved in the truncation 
which respects the  gauge identity. 
It is the  first time when the Minkowski and the Euclidean  solutions
for lowest QED Green functions were made.
Working in the Landau gauge  we have found  a good numerical agreement 
between the solutions obtained in these two technically 
different frameworks. From this we argue that the electron as well as the photon
propagator posses the standard textbook  spectral (Lehmann) representations.
We showed that at least up to the certain renormalized coupling
the proper Green functions - the photon polarization function 
as well as the electron selfenergy- satisfy appropriately subtracted  dispersion relations.
The form of them  corresponds with the results already  known 
from the conventional  perturbation theory. 
In the other words, there  is no significant signal for complex singularity of the propagators, 
which posibility is sometimes sugested in the literature. 

In the other side,
the small difference between the Minkowski and the Euclidean solutions cannot 
fully excluded such situation, but from the smallness of observed
we can speculate that this effect must be rather negligible 
even for rather strong coupling case $\alpha(p^2=0;\Lambda=10^4m)\simeq 0.4$.

Triviality of QED was confirmed in our approach. We did not find the possibility
to send the appropriate ultraviolet cutoff $\Lambda$ to infinity with simultaneous keeping 
the renormalized coupling non-zero. With increasing $\Lambda$ and 
with the non-zero bare  electron mass 
we cannot observe a second order chiral phase transition
as in the paper \cite{RAKOW} but instead of this  we observe the appearance of Landau singularity
in the running charge.
The obtained  results were also compared with bare 
vertex approximation. In that case only the small violation from our 
'gauge covariant' solution was observed. This difference is still small even for
rather strong coupling case $\alpha(p^2=0;\Lambda=10^4m)\simeq 0.5$.
The obtained solutions reduce to their perturbative counterpartners 
in small coupling limit. Furthermore, the explicit comparison with the one 
loop perturbation theory was made. Note, that the triviality statement is confirmed
in these two later cases too.

There is still missing link between theory with $m_0=0$
(in chiral symmetry phase as well as in chiral symmetry broken phase)
and our  dispersion technique in used. Nevertheless, we believe that the 
possibility to extract the
information about the timelike behavior of Green functions from
spectral approach remains the main attractive feature when 
compared with usual Euclidean approach.
After a certain automatization of the dispersion relations evaluation 
the method should be extend-able to a more
complex theories. The certain progress was already  achieved  in QCD and the results 
will be published elsewhere. There is also a broad scope for possible future
investigations in pure QED:study of $S-matrix$ property when  
it is composed from dressed Green-functions, the
study of bound states with dressed propagators, 
including transverse correction to the vertex, etc..

\begin{center}
Acknowledgments
\end{center}
This work was supported by GA \v{C}R under contract n.~202/03/0210.

%$$$$$$$$$$$$$$$$$$$$$$$$$$$$$$$$$$$$$$$$$$$$$$$$$$$$$$$$$$$$$$$$$$$$$$$$$$$$$$$
\appendix

\section{Assorted Integrals}

In this Appendix we list several useful relation. The following
integral has been used  many times in the last step of derivations of
DRs:
\be          \label{sigi}
X_n(\omega,a,b)=\int_0^1  \Theta(\omega-a/(1-x)-b/x) x^n dx \quad,
\ee
where $a,b$ are positive real numbers. For the several lowest $n$
it equals:

\bea
X_0(\omega,a,b)&=&\frac{\Delta^{1/2}(\omega,a,b)}{\omega}
\Theta(\omega-(\sqrt{a}+\sqrt{b})^2) \, ,
\nn \\
X_1(\omega,a,b)&=&\frac{\Delta^{1/2}(\omega,a,b)[\omega+b-a]}{2\omega^2}
\Theta(\omega-(\sqrt{a}+\sqrt{b})^2)  \, ,
\nn \\
X_2(\omega,a,b)&=&\frac{\Delta^{1/2}(\omega,a,b)
[(\omega+b-a)^2-\omega b]}{3\omega^3}
\Theta(\omega-(\sqrt{a}+\sqrt{b})^2)  \, ,
\nn \\
X_3(\omega,a,b)&=&\frac{\Delta^{1/2}(\omega,a,b)
[\omega+b-a][(\omega+b-a)^2-2\omega b]}{4\omega^4}
\Theta(\omega-(\sqrt{a}+\sqrt{b})^2)\, ,
\eea
where $\Delta$ is the triangle function (\ref{triangle}). The variable
$x$ in (\ref{sigi}) appears from the Feynman parametrization of
products of the inverse scalar propagators $D_{1,2}$:
\be
D_1^{-\alpha} D_2^{-\beta}=\int_0^1 dx\,
 \frac {x^{\alpha-1}(1-x)^{\beta-1}\Gamma(\alpha+\beta)}
{[D_1x+D_2(1-x)]^{ (\alpha+\beta)}\Gamma(\alpha)\Gamma(\beta)}
\ee
or from the difference of the propagators $D_{1,2}$
\be
D_1^{-\alpha}-D_2^{-\alpha}=\int_0^1 dx\, \frac {(D_2-D_1)\alpha}
{[D_1x+D_2(1-x)]^{ (\alpha+1)}}\:.
\ee
%
%$$$$$$$$$$$$$$$$$$$$$$$$$$$$$$$$$$$$$$$$$$$$$$$$$$$$$$$$$$$$$$$$$$$$
\section{ Derivation of $\Pi_{R(rem)}$}

In this Appendix we derive DR for the function $\Pi_{U(rem)}$. To this
end we formally interchange the labeling of the variables
$a\leftrightarrow b $ in the last term  of
Eq.\ (\ref{holina}). Further we substitute $l\rightarrow -l+q$ which
yields
\be
\label{domino}
\Pi_{U(rem)}(q^2)=\frac{4e^2}{3q^2}\int d_l\,
\frac{ \sigma_v(a)\sigma_s(b)\rho_S(c)\left[4l^2-16
\frac{(l\cdot q)^2}{q^2}+20l\cdot q-6q^2\right]}
{[(l-q)^2-a][l^2-b][(l-q)^2-c][l^2-c]} \, ,
\ee
where we have  used the dispersion relation formulas for $M$
(\ref{selfcon}) in order to evaluate their shifted argument difference:
\be
-\frac{M(l)-M(l-q)}{l^2-(l-q)^2}=\int dc \, \frac{\rho_S(c)}{[(l-q)^2-c][l^2-c]}
\, ,
\ee
which  is invariant under the shift $l\rightarrow -l+q$.

Next it is convenient to rewrite the product of four denominators in (\ref{domino})
making use of
\bea  \label{simply}
\left\{[(l-q)^2-a][l^2-b][(l-q)^2-c][l^2-c]\right\}^{-1}
&=&\frac{I(q;c,c)+I(q;a,b)-I(q;a,c)-I(q;c,b)}
{(c-b)(c-a)} \, , \nn
\\
I(q;a,b)&=&\left\{[(l-q)^2-a][l^2-b]\right\}^{-1}.
\eea
It is sufficient to deal only with one term on rhs. of (\ref{simply}),
the others are obtained simple by changing the spectral variables (the
logarithmic divergence appears but it cancels against the same
contribution of three remaining terms). For instance choosing the
variable $a,c$ and making a shift $x\rightarrow 1-x$ leads after the
subtraction to:
\be
\frac{4e^2}{3(4\pi)^2}\int_0^1 dx
\int_{\frac{c x+a(1-x)}{x(1-x)}}^{\infty} d\omega \,
\frac{\sigma_v(a)\sigma_s(b)(2+4x-12x^2)(q^2-\mu^2)}
{(\omega-\mu^2)(q^2-\omega)(c-b)(c-a)} \, .
\ee
Integrating over the Feynman variable $x$ yields
\be
\frac{4e^2}{3(4\pi)^2}\int_0^1 dx \int_{\frac{c x+a(1-x)}{x(1-x)}}^{\infty}
d\omega \,
\frac{ \sigma_v(a)\sigma_s(b)(2X_0+4X_1-12X_2)(q^2-\mu^2)}
{(\omega-\mu^2)(q^2-\omega)(c-b)(c-a)} \, ,
\ee
where $X_n$ is the shorthand notation for the function $
X_n(\omega,c,a)$ introduced in the Appendix A. Gathering all
expressions together one gets the dispersion relation for the
polarization function $\Pi_{R(rem)}$:
\bea  \label{dolina}
\Pi_{R(rem)}(q^2,\mu^2)&=&\frac{8e^2}{3(4\pi)^2}
\int_{m^2}^{\infty}d\omega \,
\frac{ \sigma_v(a)\sigma_s(b)(q^2-\mu^2)}
{(\omega-\mu^2)(q^2-\omega)} \nn\\
&&
\left\{\frac{F(\omega,c,c)-F(\omega,c,a)+F(\omega,b,a)-F(\omega,b,c)}{(c-b)(c-a)}\right\}
\, ,
\nn \\
F(\omega,c,a)&=&\frac{\Delta^{1/2}(\omega,c,a)}{\omega}
\left[\frac{a+c}{\omega}-2\frac{a-c}{\omega}\left(1+\frac{a-c}{\omega}\right)\right]
\Theta\left(\omega-(\sqrt{a}+\sqrt{c})^2\right)\:.
\eea

%@@@@@@@@@@@@@@@@@@@@@@@@@@@@@@@@@@@@@@@@@@@@@@@@@@@@@@@@@@@@@@@
\section{ Dispersion relation for $M_{(rem)}$}

In this Appendix we derive the absorptive part of Eq.\ (\ref{vysl}).
We start from the relation (\ref{easy}) and consider the dominant
contribution that following from the pole of photon propagator, i.e.
$\sigma_{\gamma}(b)=r_{\gamma}\delta(\beta)$. In addition we  make the
substitution $\Omega=\frac{a x+o(1-x)}{1-z}$ which leads to the
following double dispersion integral:
\bea
&&\frac{3p^2e^2}{(4\pi)^2} \int_0^1 dx \int_{ax+o(1-x)}^{\infty}
d\Omega\, \frac{ax+o(1-x)}{\Omega^2(p^2-\Omega)}\,
\frac{\sigma_v(a)\rho_S(o)}{p^2-o}
 \\
\label{vojeb}
&&\approx\frac{3p^2e^2}{2(4\pi)^2} \int_{\frac{a+o}{2}}^{\infty}
d\Omega\, \frac{a+o}{\Omega^2(p^2-\Omega)}\,
\frac{\sigma_v(a)\rho_S(o)}{p^2-o}
\eea
The last step is to use algebraic identity
\be
\frac{1}{(p^2-\Omega)(p^2-o)}=
\frac{1}{\Omega-o}\left\{\frac{1}{p^2-\Omega}-\frac{1}{p^2-o}\right\}
\label{iident} \ee
and  re-write the double DR as the difference of single DRs.
Substituting (\ref{iident}) into (\ref{vojeb}) we relabel
$\Omega\rightarrow \omega,o\rightarrow u$ in the first term and
$\Omega\rightarrow u,o\rightarrow \omega$ in the second one. This
cosmetics leads to the unsubtracted DR:
\bea
\label{pppr}
\frac{3p^2e^2}{2(4\pi)^2} \left\{P\cdot
\int_{m^2}^{\infty}du \int_{\frac{a+u}{2}}^{\infty}d\omega
\frac{(a+u)\sigma_v(a)\rho_S(u)}{\omega^2(\omega-u)(p^2-\omega)} +
P\cdot \int_{m^2}^{\infty}d\omega\int_\frac{a+\omega}{2}^{\infty}du
\frac{(\omega+a)\sigma_v(a)\rho_S(\omega)}{u^2(\omega-u)(p^2-\omega)}\right\}\:.
\eea
Taking a subtraction at the point $\mu$ we get for $M_{(rem)}(p^2)$
once subtracted DR with the  weight function $\rho_{(rem)}$:
\bea \label{kick}
\rho_{(rem)}(\omega)&=&\frac{3e^2}{2(4\pi)^2} P\cdot \int_{m^2}^{\infty}\frac{1}{\omega-u}
\nn\\
&&
\left[\frac{u+a}{\omega^2}\sigma_v(a)\rho_S(u)\Theta\left(\omega-\frac{u+a}{2}\right)+
\frac{\omega+a}{u^2}\sigma_v(a)\rho_S(\omega)\Theta\left(u-\frac{\omega+a}{2}\right)\right]
\, .
\eea

%%%%%%%%%%%%%%%%%%%%%%%%%%%%%%%%%%%%%%%%%%%%%%%%%%%%%%%%%%%%%%%%%%%%%%%%%%%%%%%%%%%
%  ______________________appended figures_________________________________________%
%%%%%%%%%%%%%%%%%%%%%%%%%%%%%%%%%%%%%%%%%%%%%%%%%%%%%%%%%%%%%%%%%%%%%%%%%%%%%%%%%%%

\newpage

\begin{figure}[t]
\label{figjedna}
\centerline{\mbox{\psfig{figure=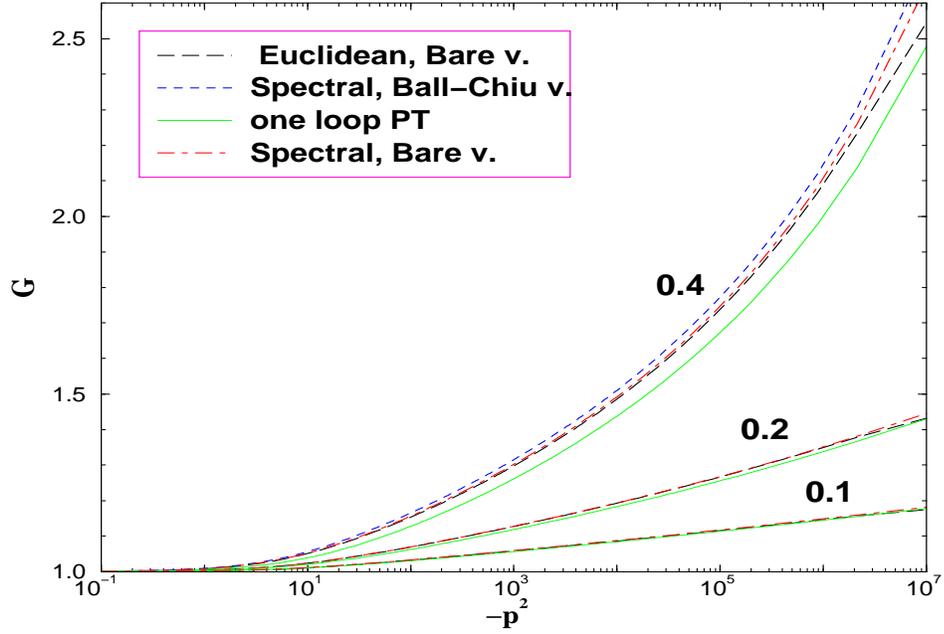,width=8.5truecm,height=12.5truecm,angle=270}
}}
\caption[caption]
{Charge renormalization function
$G=\alpha(p^2)/\alpha(0)$ obtained by solutions of DSEs. Each
beam of lines is labeled by the corresponding coupling $\alpha(0)$.
The results of leading order perturbation theory (dotted lines) always lie below
the DSE result. Only one solution with the BC is shown in the figure
(for $\alpha=0.4$), for smaller coupling the BC solutions would be undistinguishable
from the bare vertex ones.}
\end{figure}

\begin{figure}[t]
\label{figdva}
\centerline{\mbox{\psfig{figure=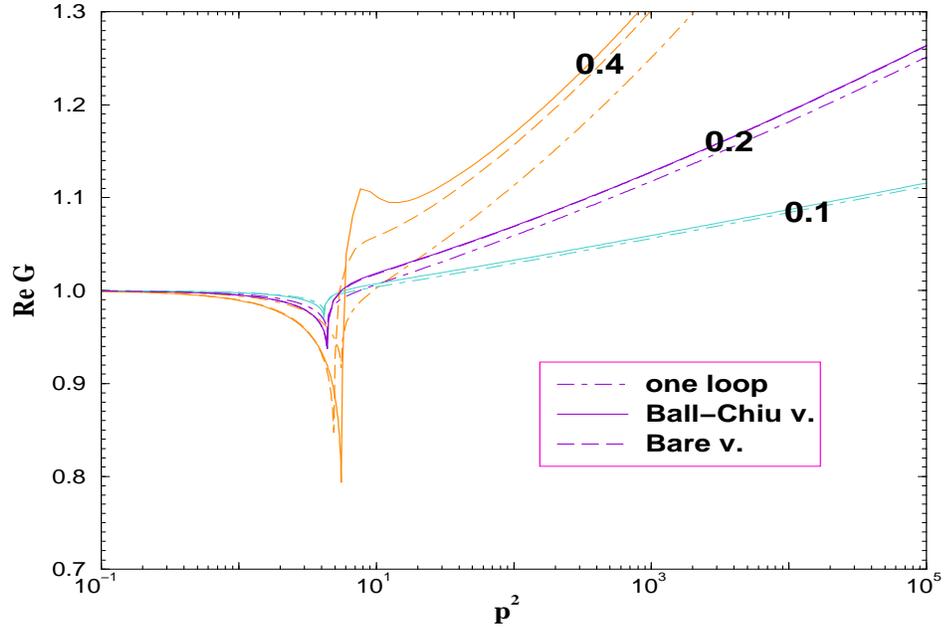,width=8.5truecm,height=12.5truecm,angle=270}
}} \caption[caption]
{Spectral solutions for charge renormalization
function $G$ for timelike momenta and for couplings:$\alpha=0.1, 0.2, 0.4$. The
down oriented peaks correspond with the threshold $4M_p^2$. }
\end{figure}

\begin{figure}[t]
\label{figtri}
\centerline{\mbox{\epsfig{figure=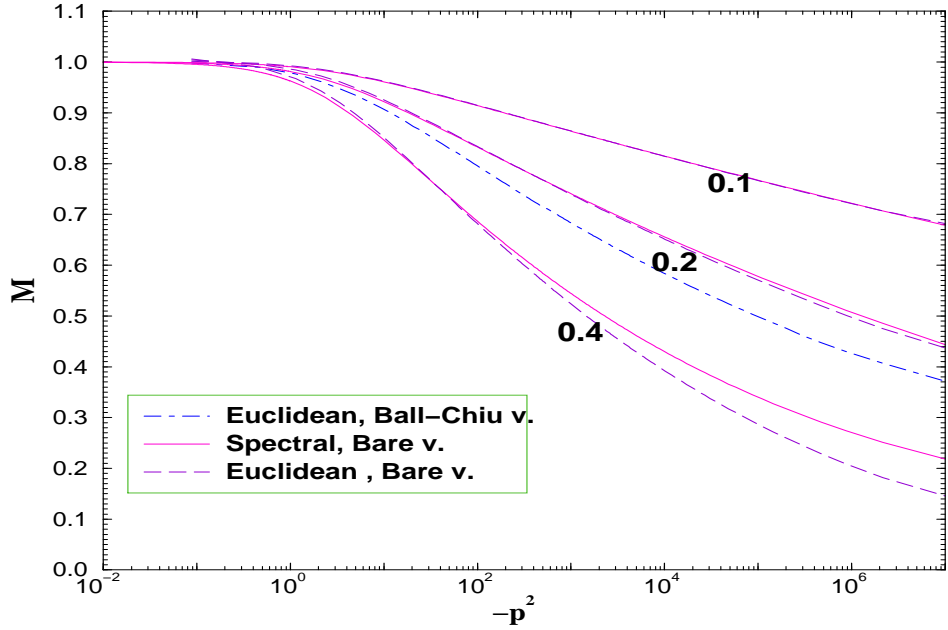,width=8.5truecm,height=12.5truecm,angle=270}
}} \caption[caption] {The comparison of dynamical mass $M(p^2)$
obtained in Euclidean and Minkowski formalism  for spacelike momenta.
For  comparison we added also one Euclidean solution calculated with
the  BC vertex (solved with approximations described in the section
IV). }
\end{figure}

\begin{figure}[t]
\label{figctyri}
\centerline{\mbox{\epsfig{figure=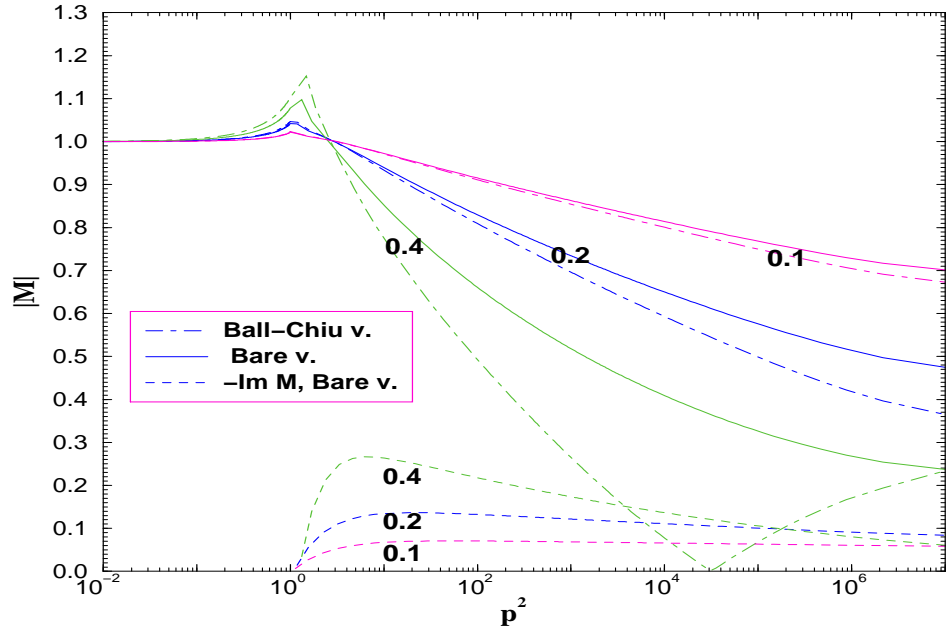,width=8.5truecm,height=12.5truecm,angle=270}
}} \caption[caption]{The absolute values of real and imaginary
parts of the electron dynamical mass as obtained by solving the
unitary  equations. The position of the up oriented peaks
correspond with the pole mass: $[M_p,M_p]$.  The  excess for the
solution with $\alpha=0.4$ is shown, the real part of $M$ becomes
negative at large momenta. }
\end{figure}

%\end{widetext}


\begin{thebibliography}{99}
\bibitem{CONV}
in this paper we use the Minkowski metric $g_{\mu\nu}=diag(1,-1,-1,-1)$, and hence
$p_E^2=-p^2$  for negative (spacelike) $p^2$.
\end{thebibliography}
\end{document}